\documentclass{aa}
\usepackage{graphicx}
\begin{document}
\thesaurus{08(08.06.2 IRAS05361+3539);09.10.1}
\title{Massive and luminous YSO IRAS 05361+3539 and its environment}
\subtitle{A study of star formation in the parent cloud - I}
\author{A. Chakraborty$^a$ 
	\inst1 
            D.K. Ojha$^b$ 
	\inst1  
            B.G. Anandarao$^c$ 
	\inst2    
	\and         
           T.N. Rengarajan
	\inst1 \thanks{\emph{Present address:} Physics Dept., Nagoya
University,  Nagoya 464-8602, Japan. e-mail: renga@u.phys.nagoya-u.ac.jp}} 
\institute{Tata Institute of Fundamental Research,
Homi Bhabha Road, Mumbai(Bombay)-400005, India  \\
e-mail: $^a$abhijit@tifr.res.in, $^b$ojha@tifr.res.in\\
\and Physical Research
Laboratory, Ahmedabad-390009, India\\
e-mail: $^c$anand@prl.ernet.in\\}
\date{}
\titlerunning{IRAS05361+3539 and its environment}
\authorrunning{Chakraborty et al.}
\maketitle
\begin{abstract}
Near-infrared photometry and narrow/broad-band imaging of the massive and
luminous young stellar object IRAS 05361+3539 are presented. Imaging
observations were made at Mt. Abu while the photometric data were taken
from the 2MASS. From the color-color and color-magnitude diagrams, we
identified several sources of faint Class II type and about six Class I type
in the parent molecular cloud complex. The IRAS 05361+3539 itself was seen to
be a Class I object and our images in Br$\gamma$ and H$_{2}$ lines show
jets/outflows from this object. The jet/outflow matches with the axis of CO
outflow detected earlier. The near-infrared and the IRAS far-infrared flux
distribution suggests a possible accretion disk with dust temperatures between
80 to 800K and extent of several tens to hundreds of AU. A possible FU Orionis
type of source was detected in the cluster.                 
\end{abstract}
\section {Introduction}
Massive YSOs are protostars (Palla \& Stahler, 1993) which are either
surrounded by ultracompact HII regions or will eventually be hot enough to produce
HII regions. They are mostly found buried deep inside clouds of gas
and dust. The major difficulties in the study of massive YSOs are (a) there
are fewer massive YSOs compared to low-mass YSOs and most of them are
at a distance greater than 1 kpc from the Sun and b) most massive YSOs suffer large
extinction (Av $\geq$ 10) and hence are difficult to study in the optical
wavelengths. However, they can be studied in the radio and infrared
wavelengths. Further, it is possible to study the environment of these objects
with
 seeing limited spatial resolution in the near-infrared using array
detectors like NICMOS. 

Recent studies by Churchwell (1997) revealed that massive YSOs
undergo similar bipolar outflows like the low-mass YSOs, but the rate of mass
outflow is larger by orders of magnitude ($\approx$3). Similarly luminosities
of massive YSOs are also higher. The total mass locked in the outflow is often
found to be larger than the central star. Churchwell (1997) proposed that the
larger outflow mass could be due to the in-falling matter directly diverted
into the bipolar jets. It is clear from earlier works (Shepherd \& Churchwell
1996; Churchwell 1997; Hartquist \& Dyson 1997; Hunter et al. 1997)
that the impact of outflows from massive YSOs shapes the history of
star-formation in the parent molecular cloud.

We have started a NIR observational program on studies of regions of massive
star formation. This is our first paper in the series. In this paper we study
the massive and luminous YSO IRAS 05361+3539 and the star formation in 
the neighborhood using 2MASS (The Two Micron All Sky Survey) data and new NIR
observations from Mt. Abu, India. The region so far is not very well studied.
IRAS source 05361+3539 (G173.58 +2.45) was studied in the millimeter lines
($^{12}$CO and $^{13}$CO) by Shepherd \& Churchwell (1996). The source is
situated at a kinematic distance of 1.8 kpc (Wouterloot \& Brand 1989). Earlier,
Wouterloot et al. (1988) detected H$_2$O maser from the source. The source is
embedded inside a large molecular cloud. The FIR fluxes meet the conditions of
Wood \& Churchwell (1989) for UCHII regions and total FIR fluxes correspond
to a B2.5 central star. Shepherd \& Churchwell (1996) found bipolar flows in
the $^{12}$CO velocity map with high
 velocities (up to -27.1 km/s in the blue
shifted lobe) with a total mass of
 32M$_{\odot}$ locked in the outflow. They
estimated from the IRAS fluxes the central star mass to be 7M$_{\odot}$. 

In section 2 we present the observations and data reduction procedures,
section 3 deals with results and discussion and we summarize our conclusions in
section 4.

\begin{table*}[ht]
\caption[]{A comparison of stellar magnitudes from 2MASS data and Mt. Abu
images. The 2MASS upper limit magnitude was derived from the histogram
of stars within the FOV of 8$^{\prime}$ towards the IRS1.}  
\[
\begin{tabular}{|c|c|c|c|c|c|c|c|}
\hline
{\it No.}&{\it Co-ordinates (J2000)}&
 \multicolumn{3}{c|}{2MASS} & 
  \multicolumn{3}{c|}{MtAbu}  \\ \cline{3-8}
%\hline
&{\it RA-DEC}&{\it Ks}&{\it H}&{\it J}&{\it K'}&{\it H}&{\it J}\\
\hline
1&05h39m22.8s +35$^{\circ}$41$^{\prime}$27$^{\prime\prime}$&
14.20&14.88&15.94&14.10&14.65&15.96\\
2&05h39m22.9s +35$^{\circ}$40$^{\prime}$22$^{\prime\prime}$&
10.08&10.11&10.40&10.09&10.11&10.35\\
3&05h39m22.9s +35$^{\circ}$41$^{\prime}$40$^{\prime\prime}$&
12.52&12.63&12.84&12.47&12.70&12.79\\
4&05h39m24.2s +35$^{\circ}$41$^{\prime}$11$^{\prime\prime}$&
12.96&13.96&16.17&12.90&13.79&16.29\\
5&05h39m24.2s +35$^{\circ}$42$^{\prime}$02$^{\prime\prime}$&
12.35&12.52&13.03&12.50&12.65&13.06\\
6&05h39m25.1s +35$^{\circ}$41$^{\prime}$12$^{\prime\prime}$&
12.22&12.27&12.54&12.33&12.39&12.52\\
7&05h39m25.5s +35$^{\circ}$40$^{\prime}$41$^{\prime\prime}$&
13.62&13.70&14.12&13.47&13.80&14.13\\
8 (IRS1)&05h39m27.0s +35$^{\circ}$40$^{\prime}$51$^{\prime\prime}$&
10.69&11.62&13.07&10.76&11.67&13.09\\
9&05h39m27.3s +35$^{\circ}$40$^{\prime}$58$^{\prime\prime}$&
12.63&13.63&15.19&12.55&13.47&15.17\\
10&05h39m27.9s +35$^{\circ}$40$^{\prime}$41$^{\prime\prime}$&
12.85&13.01&13.42&12.95&13.19&13.54\\
11&05h39m29.2s +35$^{\circ}$41$^{\prime}$39$^{\prime\prime}$&
11.57&11.69&11.88&11.80&11.85&12.06\\
12&05h39m29.3s +35$^{\circ}$41$^{\prime}$09$^{\prime\prime}$&
11.98&14.23&17.11&13.14&15.32&---\\
\hline
&Upper Limit Magnitude&15.0&15.7&16.7&14.2&15.4&16.3\\
\hline
\end{tabular}
 \]
\end{table*}                                                                    
\section {Observations and data reduction}
\subsection {Observations from Mt. Abu}
IRAS 05361+3539 (hereafter IRS1) was observed from the 1.2 meter
Infrared Telescope, Mt. Abu, India, using a 256$\times$256 HgCdTe array
(NICMOS-3, made by Infrared Laboratories, Arizona, USA). The telescope is
located at an elevation of 1700 meters from the mean sea level and the
instrument is described in Nandakumar (2000). Both the telescope and the
NIR camera are owned by the  Physical Research Laboratory, Ahmedabad, India. 

The source was observed in two sessions: in J, H and K' bands on 10th January
2000, and in narrow band filters centered on 2.12$\mu$m (H$_{2}$
$v=1-0S(1)$), 2.16$\mu$m (Br$\gamma$), and 2.14$\mu$m (continuum) on 25th
February 2000. The FWHM of the narrow-band filters was
0.042$\mu$m. The plate scale was 1$^{\prime\prime}$ / pixel for the broad-band
images and 0.5$^{\prime\prime}$/ pixel for the narrow-band images. The nights
were photometric during the observations. The seeing was 2$^{\prime\prime}$
during the broad band observations and 1.5$^{\prime\prime}$ during the narrow
band observations. A large number of dithered sky frames were obtained ( by
shifting the telescope 3 arcmin off the source in north-south-east-west
directions) in all the filters for sky subtraction and for making flat frames.
Although the 2.14$\mu$m filter is supposed to exclude the nearby emission
lines, we found from filter transmission curves that there is serious
contamination from the 2.12$\mu$m line. We therefore will not consider the
2.14$\mu$m line image in this paper and will present the narrow band images as
emission line plus continuum. We observed three standard stars (AS11, AS19,
and AS21 from Hunt et al. 1998) during the observations. We used the relation
given by   Wainscoat \& Cowie (1992) to obtain K' magnitudes from K
magnitudes. 

The data reduction was done using IRAF software tasks. All the NIR images
went through standard pipeline procedures like sky-subtraction and flat-fielding. 
Individual object frames were of 30s of integration in J and H band and 3s in
K' band. The images were co-added  to obtain a final image in each band
(J, H and K') of total integration time of 150s.  Further, in each band two
such images were obtained. The images were further filtered using a median
filter of  3$\times$3 pixels for removing noise at sky-level, thus making
them suitable for photometric analysis. We used DAOPHOT (Stetson 1987) task
for deriving the photometry in the unvignetted field of view of 2 arcmins. The
zero point was obtained using the three observed standard stars. From the J,
H and K' images containing the IRAS source, stars were identified down to
3$\sigma$ level (peak signal), subsequently each frame was visually inspected
at different contrast levels to cross check the detection and remove false
detections by DAOFIND.  Aperture photometry was performed on the images using
PHOT task with aperture radius of 4.5 pixels in the J, H and K' images. The sky
was sampled using 5-pixel wide annuli centered on each star with inner radius
at 5 pixels. The aperture size was decided using the brightest and isolated star.
From the two images of 150s in each band, we found that the overall
photometric error is of $\pm$0.07 mag. This estimation includes uncertainties
in the determination of zero-point ($\pm$ 0.03 mag) from the observed three
standard stars. We found that in the K' images we could detect 12 stars up to
14.2 mag in the field of view of 2$^{\prime}$ using the above mentioned
procedure.

The narrow-band images also went through similar image processing. The total
integration time was 250s in H$_2$ and Br$\gamma$ filters. 

\subsection {2MASS data} 
We extracted stars from the 2MASS point source catalogue which were within
8$^{\prime}$ diameter of the IRAS source. The data were downloaded from the
2MASS Homepage available free for the Astronomical community. 
The 2MASS observations were carried out on 3rd February 1998\footnote
{The date of observation was obtained from the header of the 2MASS fits images.}.
The 2MASS point
source catalogue consists of J, H and Ks magnitudes of stars. The Ks band
(bandpass =2.00$\mu$m to 2.32$\mu$m with center at 2.17$\mu$m) is very similar
to K band. A histogram of the 2MASS sources in the Ks band within a field of
view of  8$^{\prime}$ around the IRAS source was plotted. The histogram showed
that the completeness limit is close to 15 mag.

Since the 2MASS catalogue goes deeper than the Mt. Abu images we decided to
use the catalogue  magnitudes to plot color-color and color-magnitude diagram.
However, the magnitudes obtained from the Mt. Abu images will be used to find
prospective
variable stars and to study the morphology of the IRAS source and
associations.  

Table 1 shows a comparison of J,H,K' and Ks magnitudes of stars detected 
within the unvignetted field of 2$^{\prime}$ in the K' image with the 2MASS
point source catalogue. We found that for standard stars the difference in K'
and  Ks magnitudes is less than 0.05 which is less than the photometric error.
Hence we will treat the K' and Ks magnitude scales as similar in this paper.  

\section {Results and Discussion}
\begin{figure}
 \resizebox{\hsize}{!}{\includegraphics{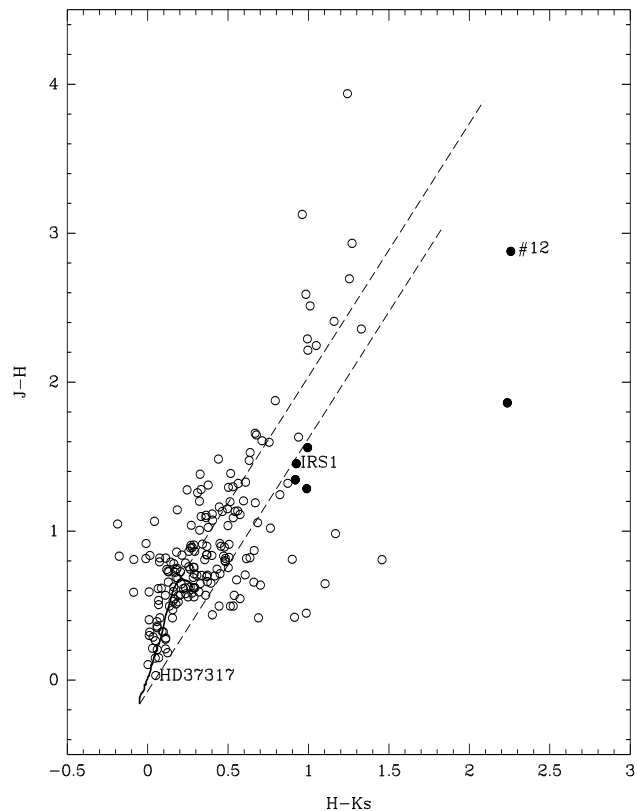}}
 \caption{Color-color diagram of the sources extracted from the 2MASS
   data. The filled circles represent Class I type sources. The solid
line represents unreddened main-sequence stars and the dashed lines are
parallel to the reddening vector with magnitude of Av=30. Also shown are the
positions of IRS1, star \# 12 (from table 1), and HD37317}    
\label{a}
\end{figure}
\begin{figure}
\resizebox{\hsize}{!}{\includegraphics{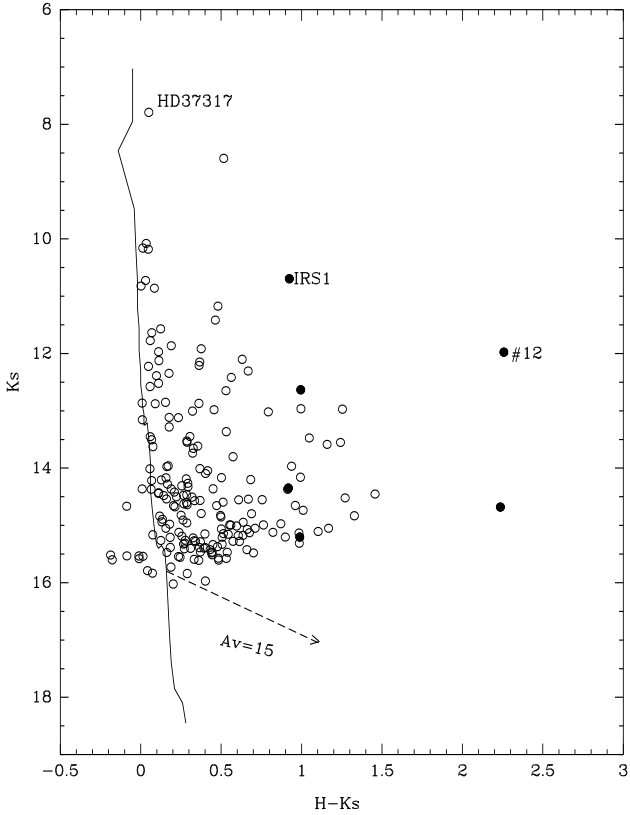}}
\caption{Color-magnitude diagram. The symbols have same meaning as in figure
1.} 
\label{b}
\end{figure}
\subsection {Star formation in the cloud}
Figures 1 and 2 show the J-H/H-Ks color-color diagram and the
Ks/H-Ks  color-magnitude diagram of the sources from 2MASS
data.  In figure 1 the solid curve is the locus of points corresponding to
unreddened main sequence stars (Koornneef, 1983). The two dashed lines are
parallel to the reddening vector with magnitude of Av=30. They form the
reddening band (drawn from the base and tip of the unreddened main sequence)
and bound the region in which stars with normal photosphere fall (also see
Hunter et al. 1995). Similarly, in Figure 2, the solid line represents the locus
of unreddened main sequence and the dashed arrow shows the direction and
magnitude of the reddening vector.  We did a search in the SIMBAD database to
find out any known sources around IRS1  with a search radius of 4.5$^{\prime}$.
Apart from the IRAS source and its H$_2$O maser source, a star of A0V type
(HD37317, 7.79 mag in Ks) was found in the SIMBAD search. This star's
co-ordinates are 05h 39m 19.9s +35$^{\circ}$ 38$^{\prime}$ 30$^{\prime\prime}$,
that is approximately 2.8$^{\prime}$ south-west from the IRAS source. We found
from the 2MASS point source catalogue the colors of the star to be
H-Ks=0.05 mag and J-H=0.031 mag. The colors of IRS1 are H-Ks=0.93 mag and
J-H=1.45 mag. 

Stars in figure 1 can be divided into three main groups a) those lying on the
left side of the reddening band, b) those lying in the reddening band and c)
those lying on the right side of the reddening band. The stars in the first
group (a) plotted as open circles can be further divided into two sub-groups:
one, those having J-H color less than 2 and second, those with J-H color more
than 2. The former subgroup are mostly foreground stars as supported by
their low values of Av (less than 5 mag), and the stars in the latter subgroup
could be spurious detections since they have J and H magnitudes fainter than 17
and 16 mag respectively. The second group (b) mostly consist of normal stars
with low values of Av ($\leq$10) and background stars with high values of Av
($\geq$10) and are also plotted as open circles. The third group (c) contains
stars showing excess emission in H and Ks. Such sources are mostly YSOs (Lada \&
Adams 1992; Lada et al. 1993; Gomez, Kenyon \& Hartmann 1994).  YSOs can be
further divided into Class I, Class II and Class III type sources based upon
their Spectral Energy Distributions (SEDs) (Strom et al. 1989; Kenyon et al.
1993; Hartmann 1998).  We have plotted sources redder than IRS1 and falling on
the right hand side of the reddening band as `filled circles' and are
considered to be Class I type or protostars (since IRS1 is a known protostar;
also see Lada \& Adams 1992). Sources which show low  H-Ks color ($\leq$1.5)
and J-H color ($\leq$1.0) and also lying on the right hand side of the strip of
reddening lines are also plotted as `open circles'. All these sources are faint
in the Ks band (14 to 15.2 mag).  Even though most of them are within the limit
of completeness, we will need deep K band images to verify the existence of the
sources. Their Av values range from 13 to 23 mag, suggesting that they could be
Class II type sources (Kenyon et al. 1993).  

It is clear from the J-H/H-Ks color-color diagram that the region around IRS1
is undergoing a phase of star formation. Three out of five Class I sources
detected (excluding  IRS1) are found within one arcmin radius of IRS1. Two of
these sources are brighter than 14.2 mag in K' and are detected in the Mt. Abu
images. A comparison of magnitudes of these two sources \#$9$ and \#$12$ (see
Table1) show that one of them \#($12$) has varied over the time of observation
between 2MASS (February 1998, as given in the header of the FITS images)  
and Mt. Abu (January 2000). The star has become fainter by 1.1 mag in K' and H
bands in about 23 months. 
In the Mt. Abu J band image the star is not detected. Since the
difference in magnitude is much larger than the photometric errors and  the
decrease in the brightness is consistent in K' and H bands, we believe that
this is real. The star also shows extreme reddening in the J-H/H-Ks color
diagram, a typical characteristic of Class I sources and presence of
circumstellar material. Variability in low-mass protostars is  known
(Hartmann 1998). The total luminosity depends on the mass accretion  rate, and
low mass protostars like FU Orionis type of objects have shown variability of
several magnitudes in the optical wavelength (Bell et al. 1995, and references
therein).  It is possible that we have witnessed a FU Orionis kind of behavior
from the star. However, we need further NIR observations to verify this.      
\begin{figure}  
\resizebox{\hsize}{!}{\includegraphics{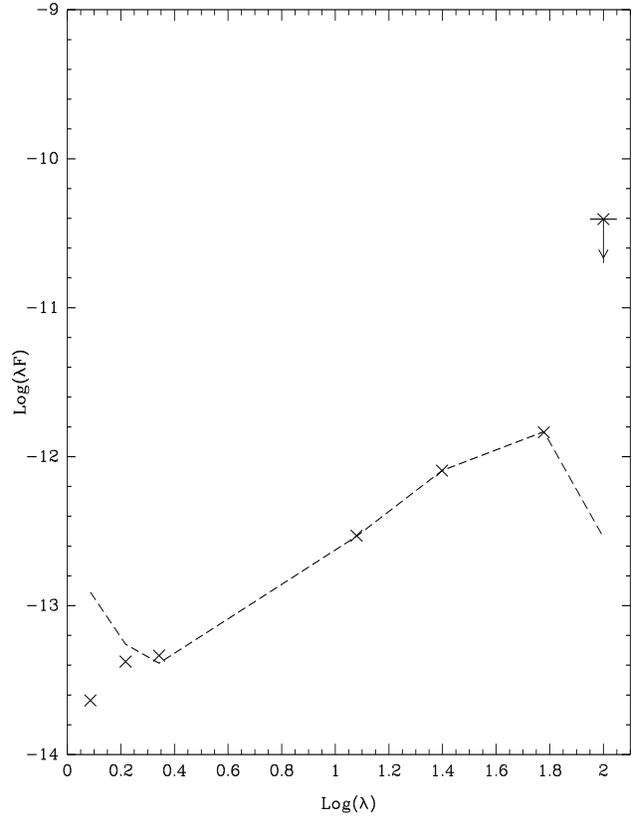}} 
\caption{SED of IRS1. Crosses represent observed fluxes and dashed line
model values. The flux F is in Watts/meter$^2$ and wavelength $\lambda$ 
in $\mu$m. See text for details.}    \label{c}
\end{figure}
\subsection{Photometry and Spectral Energy Distribution of IRAS 05361+3539}  
IRAS 05361+3539 (IRS1) is a massive luminous YSO and is associated with
an ultra-compact HII region (Shepherd \& Churchwell 1996). The low-resolution
NVSS contour plot shows emission towards IRS1 at mJy level($\leq$4)(Condon et
al. 1998). The FIR IRAS fluxes of IRS1 at 100$\mu$m, 60$\mu$m, 25$\mu$m, and
12$\mu$m are 1310Jy (upper limit), 29.15Jy, 6.72Jy, and 1.18Jy respectively. We
estimated the spectral index of IRS1 to be 1.1 between 60$\mu$m and 1.12$\mu$m
(Jband). The source therefore belongs to the group of Class I type sources
(Strom et al. 1989). The SED is shown as a plot (crosses) of Log$(\lambda
F)$/Log$(\lambda)$  in figure 3 where $\lambda$ is the wavelength in $\mu$m and
F is the flux  in Watts/meter$^2$. We determined the temperature distribution
of the circumstellar matter  by fitting a model (Anandarao, Pottasch, \& Vaidya
1993) to the observed SED. We have assumed  photospheric temperature of 20,000K,
stellar radius equal to seven solar radii, and plane parallel geometry for the
dust shells. The dashed line in Fig 3  shows the model. The derived dust
temperatures are 800K at 4AU and 80K at 400AU from the central source. The
uncertainties in the dust parameters could be as large as 10-20 \% due mainly
to the inherent non-uniqueness of the model and to some extent to the
uncertainties in the assumed stellar parameters. These results
confirm that IRS1 is a Class I type source. The range of dust shell
parameters derived from the model seem to support the accretion disk scenario
(Hartmann 1998; Adams, Lada \& Shu 1987) in which case, the grain heating is
due to two processes: one due to the reprocessing of UV photons and the other
due to viscous heating in the disk (Hillenbrand et al. 1992). Following
Hillenbrand et al. (1992) we estimate the minimum radius of an accretion disk
attributing the entire 25$\mu$m flux to the accretion heating of the grain
(i.e. T=140K). We derive radius of the disk to be 160 AU if the inclination is
0$^{\circ}$ (face on). 

\subsection{Morphology of IRAS 05361+3539 and Detection of a NIR Jet}
Figure 4 shows K', 
H$_2$ (emission line + continuum) and Br$\gamma$ (emission line +
continuum) images of IRS1. Figure 5 gives the contour map of H$_2$ superposed on the
K' image. The plate scale in Fig 4 is
1$^{\prime\prime}$/pixel for the K' image and 
0.5$^{\prime\prime}$/pixel for the narrow band images.     
\begin{figure} 
\resizebox{12cm}{!}{\includegraphics{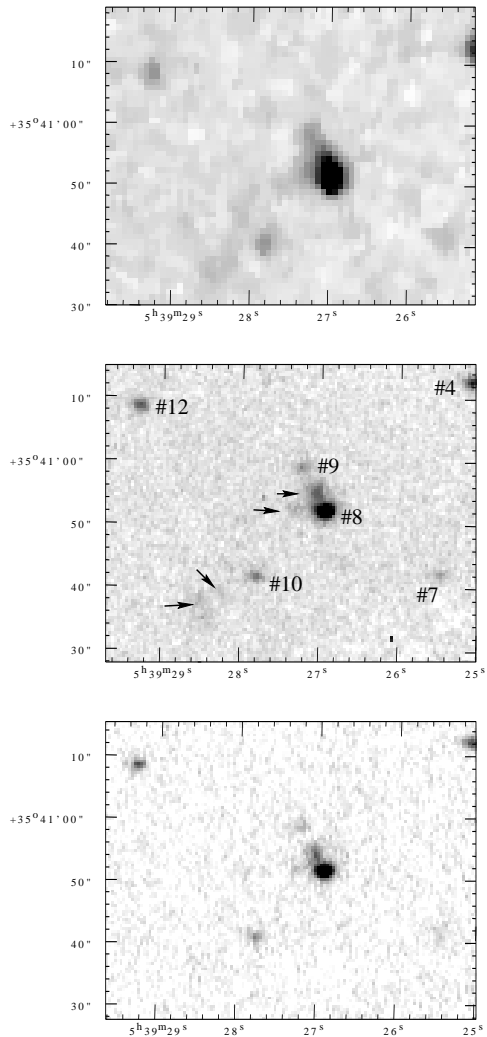}}  \hfill \parbox[b]{80mm}{
\caption{K' (top), molecular hydrogen line (2.12$\mu$m line + continuum)
(middle) and Br$\gamma$ (emission line + continuum) (bottom) images of IRS1.
The stars are identified with numbers as given in table 1 and the arrows
indicate the extended nebulosity. The x and y axes are RA and DEC (J2000)
respectively.}   \label{d}}   
\end{figure}  
\begin{figure}
\resizebox{9cm}{!}{\includegraphics{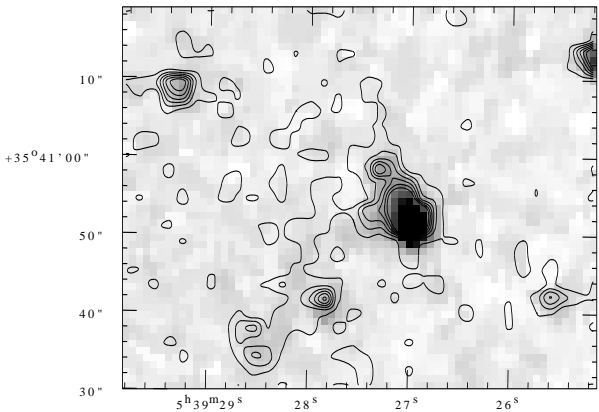}} \hfill \parbox[b]{80mm}{
\caption{Contours of the molecular hydrogen (line + continuum) image on the K'
image, the contours are from 4$\sigma$ to 9$\sigma$ levels. The x and y axes
are same as in figure 4.} \label{e}}  
\end{figure}
The NIR images reveal (figures 4, 5) that the IRS1 source is associated  
with a nebulosity extended up to 5$^{\prime\prime}$ in the northern direction 
and a filamentary structure of length 6$^{\prime\prime}$ in the east (seen more
prominently in the H$_2$+continuum image). 

The eastern filamentary structure bends beyond 7$^{\prime\prime}$ from IRS1
and continues another 20$^{\prime\prime}$ in the southeast direction. Since
we do not have a continuum free H$_2$ emission line image we cannot quantify the
amount of H$_2$ emission from the region. However, we qualitatively argue from
the brightness in the K' and the H$_2$+continuum image that at least 40\% of
the total brightness in the narrow band image is due to pure H$_2$
emission.  Further, the filamentary structure matches with the eastward 
jet found in the low resolution CO map of Shepherd \& Churchwell (1996).
Therefore, it is likely that the H$_2$ emission is tracing the jet from the YSO. The
bending of the jet could be because of physical obstruction of its flow due to the 
presence of putative dense matter. We do not see significant structure 
in the Br$\gamma$ + continuum  image. If one assumes a
particle density of 10$^5$ /cm$^3$, then there can not be sufficient flux of
energetic UV photons  available from a B2 star to ionize the 
matter except in optically thin regions.

The structure in the northern direction is six times brighter than the jet.
There are three distinct possibilities regarding the nature of this structure.
This could be an unresolved star closely associated with the IRS1 source. 
However, in the 2MASS point source catalogue there is no entry corresponding 
to that position. We will need high-resolution images in K band to prove that 
the northern structure is an unresolved star. 	

A second alternative is that the extended nebulosity is
due to the presence of the ultra-compact HII region. From the near-infrared
morphology, it appears that this could be a cometary UCHII region (Wood \&
Churchwell 1989). Supposing that the central star has a relative motion with
respect to the parent molecular cloud, say a few km/s, such a supersonic
motion can create a region of low density behind the star (Hughes \& Viner
1976; Weaver et al. 1977). If we consider a dynamical time of 5.3$\times$10$^4$
years (Shepherd \& Churchwell 1996) and a speed of 1km/s to 2km/s (Jones \&
Walker 1988) then, in as many years the star would have traveled a distance of
0.1 pc (upper limit). The extent of the extended nebulosity in the H$_2$ +
continuum image corresponds to about 0.04 pc (${\sim}$5$^{\prime\prime}$, see
figure 4). The density behind the central star should be much less (10$^3$
particles/cm$^3$) for the UV photons to reach up to 0.04 pc.

NIR spectroscopy of UCHII regions have shown that they are
bright in Br${\gamma}$ emission line (Armand et al. 1996; Doherty et al. 1994).
A$_{\gamma}$ for the UCHII region can be calculated from Av using the relation
A$_{\gamma}$/Av=0.125 (for R$_v$=5.0 extinction law of Cardelli, Clayton \&
Mathis 1989), and it turns out to be 1.87 mag. The density of matter can also be
calculated from the relation between Av and column density,
Av/N$_H$=1${\times}$ 10$^{-21}$ mag/cm$^2$ (Cardelli, Clayton \& Mathis 1989).
We determined the density of matter within 1 pc of the source assuming that most
of the extinction is from within 1pc of the source. This gives a lower limit of
density of 5x10$^3$particles/cm$^3$. The upper limit can be 10$^5$
particles/cm$^3$, if we consider that most of the extinction is within 0.05 pc. 
The value of 0.05 pc has been assumed based on the fact that very close to
the star up to 5000AU the density can be as high as 10$^8$ particles/cm$^3$ and
up to 0.5 pc it can be 10$^5$ particles/cm$^3$ (Churchwell 1997).
A sizable Stromgren sphere with the lower limit of particle density can be
obtained for a central B2 type star. But, we do not see any significant
brightness in the Br$\gamma$ + continuum image compared to the K' band image
except at the core of the nebulosity. A deeper Br$\gamma$ image will be
required for dertermining the extent of the nebulosity. 

As a third possibility, the extended nebulosity can be considered as 
a dense clump of molecular hydrogen in the parent molecular cloud. 
We found that the region is bright 
in the H$_2$+continuum image and only qualitatively we can say that pure 
H$_2$ emission is present. H$_2$ emission can arise due to either UV
fluorescence, or collisionally excited by the impact of stellar winds from
IRS1 (Genzel, 1992).

From the above arguments, it appears that it is difficult to determine 
from the present data the nature of the extended nebulosity.
High resolution radio continuum images and medium
resolution NIR spectroscopy are necessary to resolve this issue. 

\section {Conclusion}
The parent molecular cloud seems to be an active star forming region.
We have found a total of 6 prospective Class I type sources in the region
including IRAS 05361 +3539. A number of faint Class II type sources are also
detected in the region. One of the Class I sources (star no.12) detected close
to IRS1 shows extreme reddening in the J-H/H-Ks color-color diagram. This star
also appears to be fainter in H and K' band by 1.1 magnitude in the Mt. Abu
images when compared with the 2MASS magnitudes. The time difference between these
observations was 23 months. 
Therefore, star \# 12 could be a variable protostar of FU Orionis
type. We need further observations to prove it.

The infrared spectral index of IRS1 is estimated to be 1.1 
from IRAS and NIR fluxes.
Our model predicts the possibility of an accretion disk 
with dust temperatures from
80K to 800K and with an extent of several hundreds of AU. Traces of molecular
H$_2$ emissions have been detected along the CO jet axis mapped by Shepherd \&
Churchwell (1996) in the eastward direction. A nebulosity is also detected
towards north extending up to 5$^{\prime\prime}$ from IRS1. This could be due
to either an unresolved star, an UCHII region, or a dense clump of molecular
hydrogen.   

\begin{acknowledgements}
The present observations were made under a collaborative project between 
TIFR and PRL. We thank the staff at Mt. Abu IR Telescope facility 
for support during observations. We especially thank the referee Dr. Todd R.
Hunter for his very useful comments on the paper which significantly improved
the quality of the paper. This publication makes use of data products from the
Two Micron All Sky Survey, which is a joint project of the University of
Massachusetts and the Infrared Processing and Analysis Center, funded by the
National Aeronautics and Space Administration and the National Science
Foundation. This research has made use of the SIMBAD database, operated at CDS,
Strasbourg, France.   
\end{acknowledgements}


\begin{thebibliography}{}

\bibitem []{}Adams, F.C., Lada, C.J., \& Shu, F.H. 1987, ApJ, 321, 788

\bibitem []{}Anandarao, B.G., Vaidya, D.B., \& Pottasch, S.R. 1993, A\&A,
273, 570

\bibitem []{}Aramand, C., et al. 1996 A\&A, 306, 593 

\bibitem []{}Bell K.R., et al. 1995, ApJ, 444, 376 

\bibitem []{}Churchwell, E. 1997, ApJ, 479, L59

\bibitem []{}Cardelli, J.A., Clayton, G.C., \& Mathis, J.S. 1989, ApJ, 345,
245 

\bibitem []{}Condon, et al. 1998, AJ, 115, 1693

\bibitem []{}Doherty, R.M., et al. 1994, MNRAS, 266, 497

\bibitem []{}Genzel, R. 1992, The Galactic Interstellar Medium, Saas-Fee
Advenced Course 21, ed. Burton, W.B., Elmegreen, B.G., \& Genzel, R.,
(Springer-Verlag) 275

\bibitem []{}Gomez, M., Kenyon, S.J., \& Hartmann, L. 1994, AJ, 105, 1850

\bibitem []{}Hartmann, L. 1998, Accretion Processes in Star Formation,
(Cambridge University Press)

\bibitem []{}Hartquist, T.W., \& Dyson, J.E. 1997, IAU Symp. 182, Herbig-Haro
Flows and Birth of Stars, ed. B. Reipurth \& C. Bertout (Dordrecht Kluwer), 537

\bibitem []{}Hillenbrand, L.A. et al. 1992, ApJ, 397, 613

\bibitem []{}Hughes, V.A., \& Viner, M.R. 1976, ApJ, 248, 622

\bibitem []{}Hunt, L.K., et al. 1998, AJ, 115, 2594

\bibitem []{}Hunter, T.R., et al. 1997, ApJ, 478, 283

\bibitem []{}Hunter, T.R., et al. 1995, A\&A, 302, 249

\bibitem []{}Jones, B.F., \& Walker, M.F. 1988 AJ, 95, 1755 

\bibitem []{}Kenyon, S.J., et al. 1993, ApJ, 414, 773

\bibitem []{}Koornneef, J. 1983, A\&A, 128, 84

\bibitem []{}Lada, C.J., \& Adams, F.C. 1992, ApJ, 393, 278

\bibitem []{}Lada, C.J., et al. 1993, ApJ, 408, 471

\bibitem []{}Nandakumar, M.S.  2000, Ph.D. Thesis, Gujarat University,
Ahmedabad 

\bibitem []{}Palla, F., \& Stahler, S.W. 1993, ApJ, 418, 414

\bibitem []{}Shepherd, D.S., \& Churchwell, E. 1996, ApJ, 472, 225

\bibitem []{}Stetson, P.B. 1987, PASP, 99, 191

\bibitem []{}Strom, K.R., et al. 1989, ApJS, 71, 183

\bibitem []{}Wainscoat, R.J., \& Cowie, L.L. 1992, AJ, 103, 332

\bibitem []{}Weaver, R., et al. 1977, ApJ, 218, 377

\bibitem []{}Wood, D.O.S., \& Churchwell, E. 1989, ApJS, 69, 831

\bibitem []{}Wouterloot, J.G.A., \& Brand, J. 1989, A\&AS, 80,149

\bibitem []{}Wouterloot, J.G.A., et al. 1988, A\&A, 191, 323                 
                 
\end{thebibliography}
\end{document}